# PROJECT X – A NEW MULTI-MEGAWATT PROTON SOURCE AT FERMILAB*

S. Nagaitsev, Fermilab, Batavia, IL, 60510, U.S.A.

*Abstract*

Project X is a multi-megawatt proton facility being developed to support intensity frontier research in elementary particle physics, with possible applications to nuclear physics and nuclear energy research, at Fermilab. The centerpiece of this program is a superconducting H- linac that will support world leading programs in long baseline neutrino experimentation and the study of rare processes. Based on technology shared with the International Linear Collider (ILC), Project X will provide multi-MW beams at 60-120 GeV from the Main Injector, simultaneous with very high intensity beams at lower energies. Project X will also support development of a Muon Collider as a future facility at the energy frontier.

## PROJECT X MISSION AND GOALS

Project X is a high intensity proton facility being developed to support a world-leading U.S. program in Intensity Frontier over the next several decades. Project X is an integral part of the long range strategic plan for the U.S. Department of Energy (DOE) High Energy Physics program as described in the P5 report of May 2008 [1], and the Fermilab Steering Group Report of August 2007 [2].

The primary mission elements to be supported by Project X include:

1. Prove a neutrino beam for long baseline neutrino oscillation experiments, based on a capability of targeting at least 2 MW of proton beam power at any energy between 60 – 120 GeV.
2. Provide MW-class, multi-GeV, proton beams supporting multiple kaon, muon, and neutrino based precision experiments. Simultaneous operations of the rare processes and neutrino programs are required.
3. Provide a path toward a muon source for a possible future Neutrino Factory and/or a Muon Collider.
4. Provide options for implementing a program of Standard Model tests with nuclei and/or nuclear energy applications

These elements represent the fundamental design criteria for Project X.

The development of a design concept for a high intensity proton facility has gone through several iterations culminating in a concept, designated the Project X Reference Design [3], that meets the high level design criteria listed above in an innovative and flexible manner. The Reference Design is based on a 3 GeV superconducting CW linac, augmented by a superconducting pulsed linac for acceleration from 3-8 GeV, and modifications to the existing Recycler and Main Injector Rings at Fermilab. The Reference Design provides a facility that will be unique in the world with unmatched capabilities for the delivery of very high beam power with flexible beam formats to multiple users.

## PROJECT X REFERENCE DESIGN

Figure 1 presents a schematic depiction of the Project X Reference Design aligned with the mission elements outlined in Section I. The primary elements are:

- An H- source consisting of a 2.5 MeV RFQ, and Medium Energy Beam Transport (MEBT) augmented with a wideband chopper capable of accepting or rejecting bunches in arbitrary patterns at up to 162.5 MHz;
- A 3 GeV superconducting linac operating in CW mode, and capable of accelerating an average (averaged over >1 μsec) beam current of 1mA, and a peak beam current (averaged over <1 μsec) of 10 mA;
- A 3 to 8 GeV pulsed superconducting linac capable of accelerating 1-mA beam pulses with a 1-5% duty cycle;
- A pulsed dipole that can split the 3 GeV beam between the neutrino program and the rare processes program;
- An rf beam splitter that can deliver the 3 GeV beam to multiple (at least three) experimental areas;
- Experimental facilities and detectors to support an initial round of 3 GeV experiments;
- Modification to the Recycler and Main Injector Ring required to support delivery of 2 MW of beam power from the Main Injector at any energy between 60-120 GeV;
- All interconnecting beam transport lines.

To guide the design and to support the mission elements a set of Functional Requirements have been established for Project X [4].

## PROJECT X LINACS

The Project X 3-GeV CW linac is being designed to accelerate an H-/p beam with an average current of 1 mA with the maximum bunch repetition rate of 162.5 MHz. However, the bunch pattern (timing) requirements vary depending on the experimental program. Fortunately, the superconductive cavities store high enough rf field energy to support short bursts of higher beam current, while maintaining an average current of 1 mA over a period of 1-2 μs. Such a period between beam bursts is well matched for a program requiring slow or stopped muons. It is envisioned that the ion source and the RFQ would be capable of accelerating beam currents of up to 5 mA. The selective removal of bunches to reduce the average beam current to 1 mA would be accomplished by a flexible broad-band chopper [5]. There are two principal timelines associated with beam chopping: (1) the timeline for strip injection into the Recycler/MI and (2) the timeline for the 3-GeV program. The injection into the ring requires 2-5% of the linac duty cycle. The total charge needed for injection is 26 mA-ms every 0.7-1.4 seconds (determined by the MI ramp cycle). The required bunch chopping during this timeline is associated with the ring rf frequency (~53 MHz) and with the kicker gap needed every revolution period (11 μs). The remainder of the duty cycle (>95%) is delivered to the 3-GeV experiments. The timeline for this program is repeated every 1-2 μs and is designed to provide different bunch structures <u>simultaneously</u> to several different experiments. Figure 2 illustrates the functioning of the chopper and the deflection cavity to provide tailored bunch patterns to three experiments. The nominal bunch repetition rate after the RFQ is 162.5 MHz. However, some bunches are removed in a manner that provides a 100 nsec burst out of each 1 μs ("red" bunches), while keeping "green" and "blue" bunches at a constant rate of 10 and 20 MHz. The average current is maintained at 1 mA while the ion source is providing 4.2 mA with the ~75% of bunches removed by the chopper. The deflecting-mode rf cavity operating at n±1/4 of the nominal bunch frequency separates the bunches at 3 GeV to three different experiments.

---

* Work supported by the Fermi Research Alliance, under contract De-AC02-07CH11359 to the U.S. Department of Energy

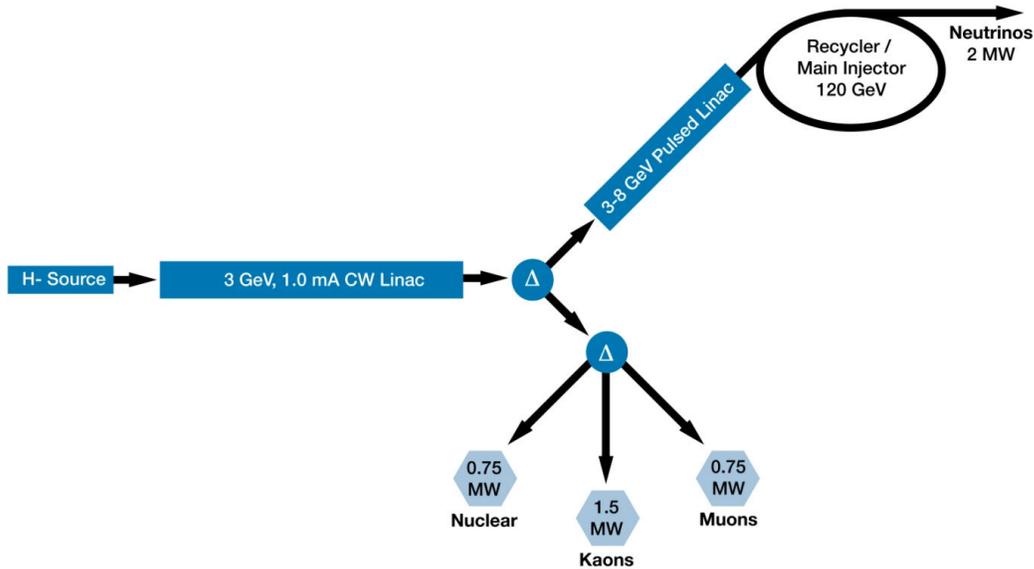

Figure 1: Project X Reference Design

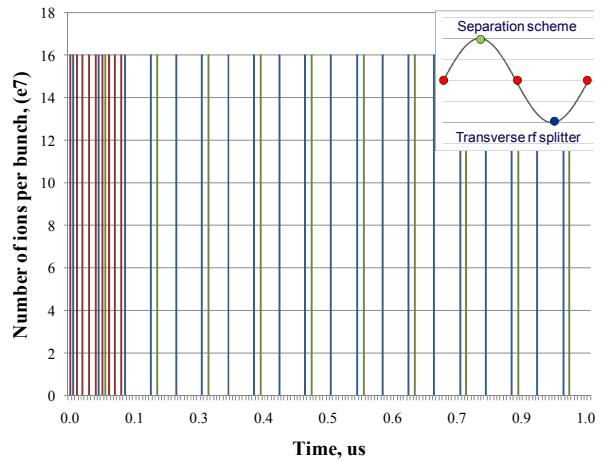

Figure 2: Illustration of the utilization of the 2.5 meV chopper and the 3-GeV bunch separator to provide a particular bunch pattern corresponding to beam powers in Fig.1. Inset: The arrival time times of various bunches at the deflecting-mode separator cavity (red –muons, blue – kaons, and green – nuclear)

The entire linac downstream of the ion soure, RFQ and the MEBT/chopper section is based on superconductive cavities. Figure 3 shows the deployment of these cavities. A total of six cavity types at three different frequencies are utilized. The 325-MHz accelerating structures are of the single-spoke resonator type, optimized for β's of 0.11, 0.22, and 0.4 respectively. The 650-MHz structures are of (5-cell) elliptical shape optimized for β's of 0.6 and 0.9. The 1.3-MHz structures are TESLA-type cavities of the ILC configuration, with suitable modifications for a long-pulse operation.

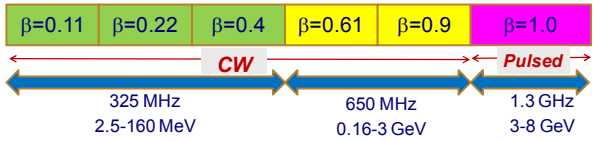

Figure 3: A block diagram of Project X linac cavity types.

Approximately 280 cavities of various types are required for the cw linac and 224 cavities for the pulsed linac. The physics design of the linac is presented in Ref. [6] and [7].

## MAIN INJECTOR AND RECYCLER

The Project X Reference Design assumes that the H- ions would be strip-injected into the Recycler ring for 4.3 ms at a 10-Hz rate. The Recycler will operate as an accumulation ring, using boxcar stacking of 6 pulses from the linac, capturing in 53 MHz RF buckets, and performing a single turn extraction into the MI. The MI will receive $1.6 \times 10^{14}$ protons from the Recycler in a single turn and will accelerate them to 120 GeV in 1.2 seconds. The Fermilab Recycler Ring is a fixed energy 8 GeV storage ring using strontium ferrite permanent magnets in the Main Injector (MI) tunnel. For the NOvA program, the Recycler will be converted from an antiproton storage ring to a proton accumulator for single turn injection into the Main Injector. In general, the Recycler and Main Injector operate in a similar mode to the NOvA operation.

The Reference Design calls for a thin carbon foil as the default stripping system at injection. We also plan on investigating the laser stripping process that has been demonstrated at SNS for 1 GeV H-.

The injection process itself consists of both transverse and longitudinal phase space painting to create a "KV-like" distribution and minimize space-charge tune shift in the Recycler. A combination of horizontal painting and vertical injection steering is used to minimize the required vertical aperture and reduce the complexity of painting in the ring in both dimensions to produce a uniform transverse distribution in x and y.

In the present Main Injector, the injection energy is 8 GeV and maximum energy 150 GeV. Transition crossing is at $\gamma_t$ = 21.6 GeV. Simulation shows that emittance dilution and beam loss will occur at high intensities. Furthermore, from experience at other machines (e.g., the AGS at BNL, the CERN PS, and the KEK PS), transition crossing could become a severe bottleneck in high intensity operation. The design being considered is a first-order system employing local dispersion inserts at dispersion-free straight sections. The normal ramp rate of the MI is 240 GeV/s. In order to have an effective $\gamma_t$-jump, the jump rate should be at least an order of magnitude higher. The system was chosen to provide a $\Delta\gamma_t$ from +1 to -1 within 0.5 ms, a jump rate of 4000 GeV/s, about 17 times faster than the normal ramp rate. Details can be found in the Proton Driver Design study document [8].

Electron cloud induced instabilities in either the Recycler or the MI could be an important limitation to the maximum proton flux. Ongoing studies in the Main Injector are investigating the generation of electron clouds and the performance of beam pipe coatings (e.g., TiN) in reducing the secondary electron yield. Simulations are being benchmarked to measurements. Based on the measurements and simulations, it appears that a combination of beam pipe coating and instability damper upgrades will be enough to mitigate electron cloud instabilities.

Upgraded RF systems are required in both the Recycler and the Main Injector. The Recycler RF system is used solely for capture of the injected beam. With the higher intensity, the Main Injector needs more power than is currently available to accelerate the beam at 240 GeV/s. To minimize changes to the existing instrumentation (specifically BPM systems), the RF will operate at the same harmonic number of 588.

The peak beam current in the Main Injector is 2.25 A and 2.7 MV/turn are necessary to reach the desired acceleration rate. With a synchronous phase angle $\phi_s$ = 36˚, 240 kV per cavity are required. From injection energy (8 GeV) to flattop (120 GeV), the frequency sweep is from 52.811 MHz to 53.104 MHz. A cavity design is under development, to be used in both the Recycler and the Main Injector.

To mitigate possible space charge effects, an increased bunching factor is desired. A frequency mismatch between the capture RF and the linac RF induces a parasitic longitudinal painting but that is not sufficient. Studies have shown that a 2nd RF system operating at the 2nd harmonic and at ½ the voltage increase the bunching factor to ~0.35, within the desired range to mitigate space charge effects.

## BEAM TRANSPORT

The H- transport should have sufficiently small loss to minimize residual radiation in the tunnel. It is highly desirable to keep residual radiation level well below 20 mRem/hr. Many facilities use the metric of 1 W/m as a limit for "hands on" maintenance, however, at 3 GeV, a 1W/m loss produces a peak contact residual dose rate of ~150 mRem/hr on a bare beam pipe. This loss rate produces significantly lower residual activation on external surfaces of magnets due to shielding by the lamination steel. However, a main concern is the residual level at magnet interfaces and instrumentation locations. These levels are based upon MARS estimations and used for order of magnitude estimations. A more accurate estimation will be required once a detailed model of the transport line is available. Setting a desirable activation level to 10 mRem/hr results in a loss goal of ~0.1 W/m.

The single particle mechanisms contributing to the beam loss are the Lorentz stripping in dipoles, the beam stripping in the residual gas, the photo detachment by blackbody radiation in the beam pipe, and loss due to intra-beam stripping. All four mechanisms are important for the CW beam transport at 3 GeV; at 8 GeV the blackbody radiation and the Lorentz stripping dominate. Simulations indicate that the loss rate goals can be achieved by keeping the magnetic field magnitude below 1.2 kG at 3 GeV and below 0.5 kG at 8 GeV, and by lowering the beam pipe temperature to about 150 K.

## PROJECT X AND MUON FACILITIES

Project X shares many features in common with the proton driver required for a muon-based accelerator facility – either a Neutrino Factory or a Muon Collider. Both requie protons directed onto a production target at an energy between 5 and 15 GeV, with a total beam power of approximately 4 MW. This power is well within the upgrade capabilities of Project X. However, the beam delivered from the CW linac (or from the pulsed linac, for that matter) does not carry the correct beam format. The muon facilities generally require the proton beam consolidated into a few very short bunches (~2 ns rms), repeating at tens of Hz. It appears inevitable that at least two new rings (for accumulation and bunch compression) would be required downstream of Project X to provide the required bunch format. A conceptual design of these two rings will be developed over the next several years.

## SUMMARY

Project X is central to Fermilab's strategy for future development of the accelerator complex. Project X will support a world leading program in neutrinos and other rare processes over the coming decades, and will be constructed in a manner that could provide a stepping stone to muon-based facilities – either a Neutrino Factory or a Muon Collider. A reference design has been completed which satisfies the four primary mission elements and provides a flexible platform for future development of the Fermilab accelerator complex.

The current configuration supports in excess of 2 MW of beam power at any energy between 60-120 GeV, simultaneous with 3MW at 3 GeV. Multiple experiments can be supported with varying beam requirements. The CW linac is unique within the world and offers capabilities that will be very difficult to duplicate in a synchrotron.

A multi-institutional collaboration has been formed to undertake the Project X R&D program with Fermilab as the lead laboratory. While the bulk of R&D will be concentrated in the area of superconductive rf, there are other outstanding technical issues being pursued, specifically, 1) the broad-band chopper, 2) the multi-turn stripping injection and 3) electron-cloud effects.

With adequate support Project X could be constructed over the period 2015-2019, providing a unique facility for physics research starting around 2020.

## REFERENCES

[1] "U.S. Particle Physics: Scientific Opportunities, A Strategic Plan for the Next Ten Years", May 2008, http://www.science.doe.gov/hep/files/pdfs/P5_Report%2006022008.pdf
[2] Fermilab Steering Group Report, September 2007, http://www.fnal.gov/pub/directorate/steering/index.shtml
[3] Project X Reference Design Report, November 2010, http://projectx-docdb.fnal.gov:8080/cgi-bin/ShowDocument?docid=776
[4] S.D. Holmes, et al., "Project X Functional Requirements Specification", these proceedings.
[5] V. Lebedev, et al., "Broad-band Beam Chopper for a cw Proton Linac at Fermilab", these proceedings.
[6] N. Solyak, et al., "Physics Design of the Project X CW Linac", these proceedings.
[7] N. Solyak, et al., "Conceptual Design of the Project-X 1.3-GHz 3-8 GeV Pulsed Linac", these proceedings.
[8] G.W.Foster, W. Chou, and E. Malamud (ed), "Proton Driver Study II", Fermilab-TM-2169, (May 2002). The γ-t jump system is described in Chapter 16.